\documentclass[conference]{IEEEtran}
\IEEEoverridecommandlockouts
\usepackage{cite}
\usepackage{amsmath,amssymb,amsfonts}
\usepackage{algorithmic}
\usepackage{graphicx}
\usepackage{textcomp}
\usepackage{xcolor}

\usepackage{algorithm}
\usepackage{algorithmic}
\usepackage{pdfpages}
\usepackage{multirow}
\usepackage[export]{adjustbox} 
\usepackage{bm}
\usepackage{amsmath}
\usepackage{array}
\usepackage{booktabs}
\usepackage{footmisc}

\usepackage{amsthm}
\usepackage{graphicx}
\usepackage{enumitem}

\theoremstyle{Definition}

\usepackage{subfigure}
\usepackage{hyperref}
\usepackage{url}
\usepackage{color}
\usepackage{xcolor}
\usepackage{colortbl}
\usepackage{xspace}
\newcommand{\model}{IMGCF\xspace}

\usepackage{amsmath,amsfonts,bm}

\def\eqref#1{equation~\ref{#1}}

\def\1{\bm{1}}

\DeclareMathAlphabet{\mathsfit}{\encodingdefault}{\sfdefault}{m}{sl}
\SetMathAlphabet{\mathsfit}{bold}{\encodingdefault}{\sfdefault}{bx}{n}

\def\BibTeX{{\rm B\kern-.05em{\sc i\kern-.025em b}\kern-.08em
    T\kern-.1667em\lower.7ex\hbox{E}\kern-.125emX}}

\makeatletter
\newcommand{\linebreakand}{%
    \end{@IEEEauthorhalign}
    \hfill\mbox{}\par
    \mbox{}\hfill\begin{@IEEEauthorhalign}}
\makeatother

\begin{document}

\title{Alleviating Behavior Data Imbalance for Multi-Behavior Graph Collaborative Filtering}


\author{
\IEEEauthorblockN{Yijie Zhang$^1$, Yuanchen Bei$^2$, Shiqi Yang$^1$, Hao Chen$^3$,}
\IEEEauthorblockN{Zhiqing Li$^4$, Lijia Chen$^5$, Feiran Huang$^1$}
\IEEEauthorblockA{$^1$ \textit{Jinan University, Guangzhou, China}}
\IEEEauthorblockA{$^2$ \textit{Zhejiang University, Hangzhou, China}}
\IEEEauthorblockA{$^3$ \textit{The Hong Kong Polytechnic University, Hong Kong SAR, China}}
\IEEEauthorblockA{$^4$ \textit{South China Agricultural University, Guangzhou, China}}
\IEEEauthorblockA{$^5$ \textit{Guangdong University of Finance, Guangzhou, China}}
\IEEEauthorblockA{wingszhangyijie@gmail.com,
iyuanchenbei@gmail.com,
mollyshiqiyang@gmail.com,
sundaychenhao@gmail.com,}
\IEEEauthorblockA{zhiqinglizzy@gmail.com,
lijarachen@gmail.com,
huangfr@jnu.edu.cn}
}

\maketitle

\begin{abstract}
Graph collaborative filtering, which learns user and item representations through message propagation over the user-item interaction graph, has been shown to effectively enhance recommendation performance.
However, most current graph collaborative filtering models mainly construct the interaction graph on a single behavior domain (e.g. click), even though users exhibit various types of behaviors on real-world platforms, including actions like \textit{click}, \textit{cart}, and \textit{purchase}.
Furthermore, due to variations in user engagement, there exists an imbalance in the scale of different types of behaviors. For instance, users may click and view multiple items but only make selective purchases from a small subset of them.
How to alleviate the behavior imbalance problem and utilize information from the multiple behavior graphs concurrently to improve the target behavior conversion (e.g. purchase) remains underexplored.
To this end, we propose \model, a simple but effective model to alleviate behavior data imbalance for multi-behavior graph collaborative filtering. Specifically, \model utilizes a multi-task learning framework for collaborative filtering on multi-behavior graphs. Then, to mitigate the data imbalance issue, \model improves representation learning on the sparse behavior by leveraging representations learned from the behavior domain with abundant data volumes. Experiments on two widely-used multi-behavior datasets demonstrate the effectiveness of \model.
\end{abstract}

\begin{IEEEkeywords}
recommender systems, multi-behavior recommendation, graph neural network, multi-task learning
\end{IEEEkeywords}

\section{Introduction}
With the information explosion, recommender systems have become widely deployed in real-world applications to effectively curate relevant information to help users discover items of interest from the overwhelming volume of available data~\cite{wu2022survey,gao2023survey,bei2023non}.
Hence, accurately modeling user interests is a core capability of recommendation systems.
As an effective technique, collaborative filtering (CF)~\cite{su2009survey} that learns from historical user-item interactions, has proven to be particularly successful in this regard~\cite{he2017neural,wang2019neural,xu2022flattened,huang2023align}.

Learning informative representations of users and items plays a pivotal role in enhancing CF effectiveness.
Early works like matrix factorization (MF)-based models individually project the user/item ID into the representation vectors~\cite{koren2009matrix,xue2017deep}, largely disregarding the high-order user-item interactions.
Recently, the graph neural network (GNN)-based models organize the historical user-item interactions as a bipartite graph and apply message propagation over the interaction graph for collaborative filtering~\cite{kipf2016semi,wu2020comprehensive}, which allows for more nuanced modeling of high-order user-item relationships and achieves the state-of-the-art performance~\cite{wang2019neural,he2020lightgcn}.

\begin{figure}[tbp]
    \centering
    \includegraphics[width=\linewidth]{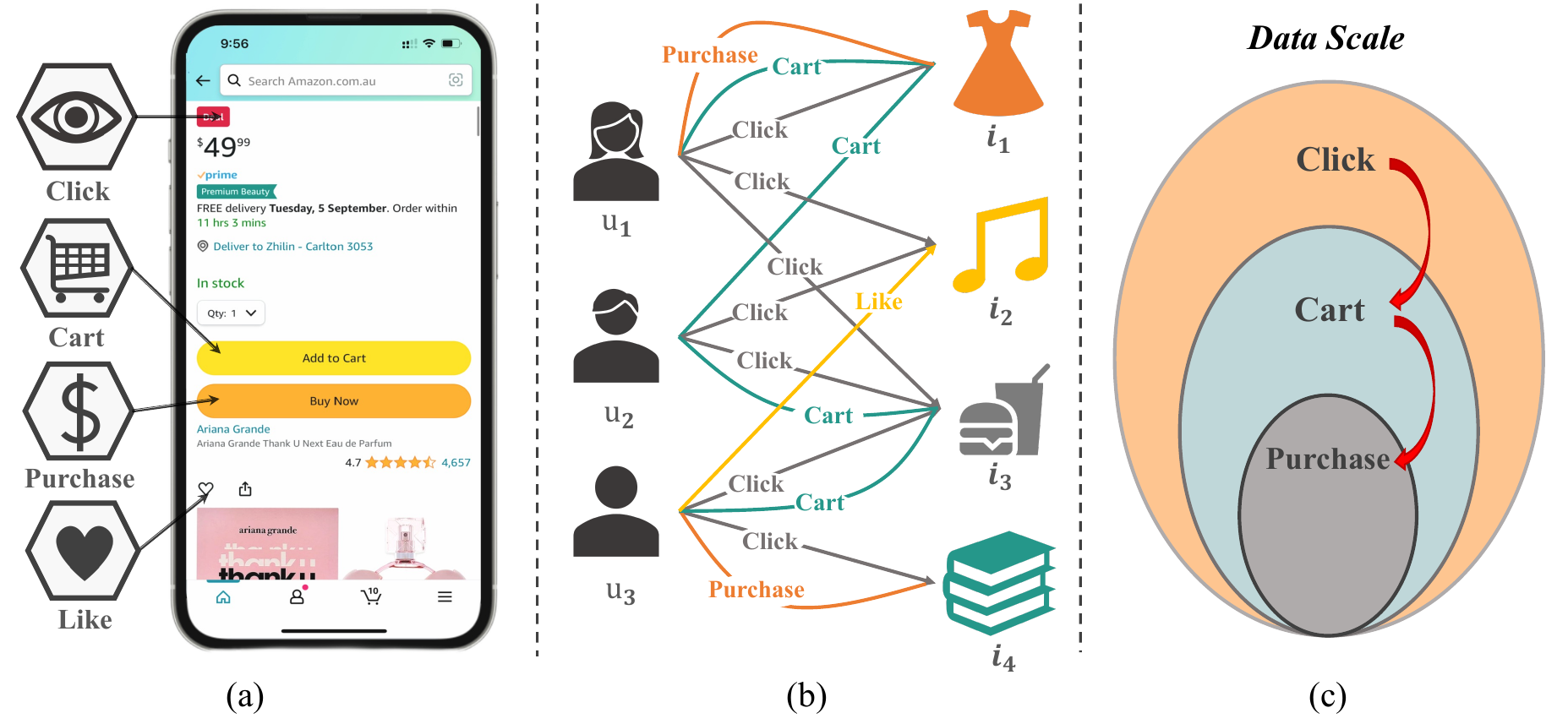}
    \caption{An illustration of the user's multiple behaviors in a real-world platform and the multi-behavior data scale imbalance phenomenon.}
    \label{fig:user-behavior}
\end{figure}

Despite the success of these graph collaborative filtering models in learning from user behavior data, there remains a gap in their practical application for real-world recommendations for the following challenges: 
(i) \textbf{Multiple types of behaviors co-occurrence}: As shown in Figure \ref{fig:user-behavior}-(a) and \ref{fig:user-behavior}-(b), in the real-world platform, users may have multiple types of behaviors, such as \textit{click}, \textit{cart}, and \textit{purchase}. Understanding this multidimensional user behavior can significantly enhance the precision of recommendations~\cite{jin2020multi,xia2021knowledge,xia2021multi}.
However, existing graph collaborative filtering methods typically focus on a single domain of behavior graph (e.g. click) for user modeling, neglecting the potential benefits of multi-behavior modeling for enhancing target behavior conversion.
Recently, some works have begun to study multi-behavior modeling and achieved successful improvements~\cite{ma2018entire,ma2018modeling,tang2020progressive}. Nevertheless, how to concurrently incorporate multiple types of behaviors into graph collaborative filtering remains largely underexplored.
(ii) \textbf{Behavior data imbalance}: As illustrated in Figure~\ref{fig:user-behavior}-(c), there is a significant imbalance in the quantity of data among different types of behaviors. Users may click on many items, but only add a portion to the cart, and ultimately purchase only a few satisfactory items.
How to alleviate such an imbalance problem for improving the modeling performance of the target behavior remains a barrier.

Therefore, to address the above issues, we propose \textbf{\model}, a novel method to alleviate behavior data \underline{I}mbalance for \underline{M}ulti-behavior \underline{G}raph \underline{C}ollaborative \underline{F}iltering. 
Specifically, \model first treats learning on each behavior domain as a task and applies a multi-task learning architecture for multiple behavior graph modeling.
Then, to leverage the rich information from abundant behavior types, we enhance graph representation learning on sparse behaviors by aggregating the representations learned from these rich behavior types.
Finally, \model jointly optimizes the multi-task graph collaborative filtering to improve the sparse target behavior conversion performance.
Experimental results on two benchmark datasets demonstrate the effectiveness of \model.
The main contributions of this paper are summarized as follows:
\begin{itemize}
    \item We highlight the challenges of utilizing graph collaborative filtering over multi-behavior graphs with behavior data imbalance in the quantity.
    \item We propose \model, a novel method to alleviate the behavior data imbalance problem in multi-behavior graph collaborative filtering.
    \item We conduct extensive experiments on two multi-behavior benchmark datasets. The results show that  \model makes state-of-the-art performance on the target behavior.
\end{itemize}

\section{Preliminaries}
\subsection{Multi-Behavior Recommendation}
Multi-behavior recommendation leverages various user-item behavioral relationships to enhance recommendations~\cite{xia2021multi,zhang2023denoising,cheng2023multi}. Let $\mathcal{U}$ and $\mathcal{I}$ denote the user set and item set, and the number of set elements is recorded as $|\mathcal{U}|=M, |\mathcal{I}|=N$. We define $\bm{R}_{u,i}^k=1$ to denote interaction between the user $u$ and item $i$ in the $k$-th behavior domain, otherwise $\bm{R}_{u,i}^k=0$. Then, the interaction matrices of $K$ behaviors are $[\bm{R}^1, \dots, \bm{R}^k, \dots, \bm{R}^K]$.
In general, the objective of multi-behavior recommendation is to utilize information from other behaviors, like clicking and carting behaviors, which are called source behaviors, to provide recommendations to users in the target behavior domain, such as purchasing.


\begin{figure*}[t]
    \centering
    \includegraphics[width=\linewidth]{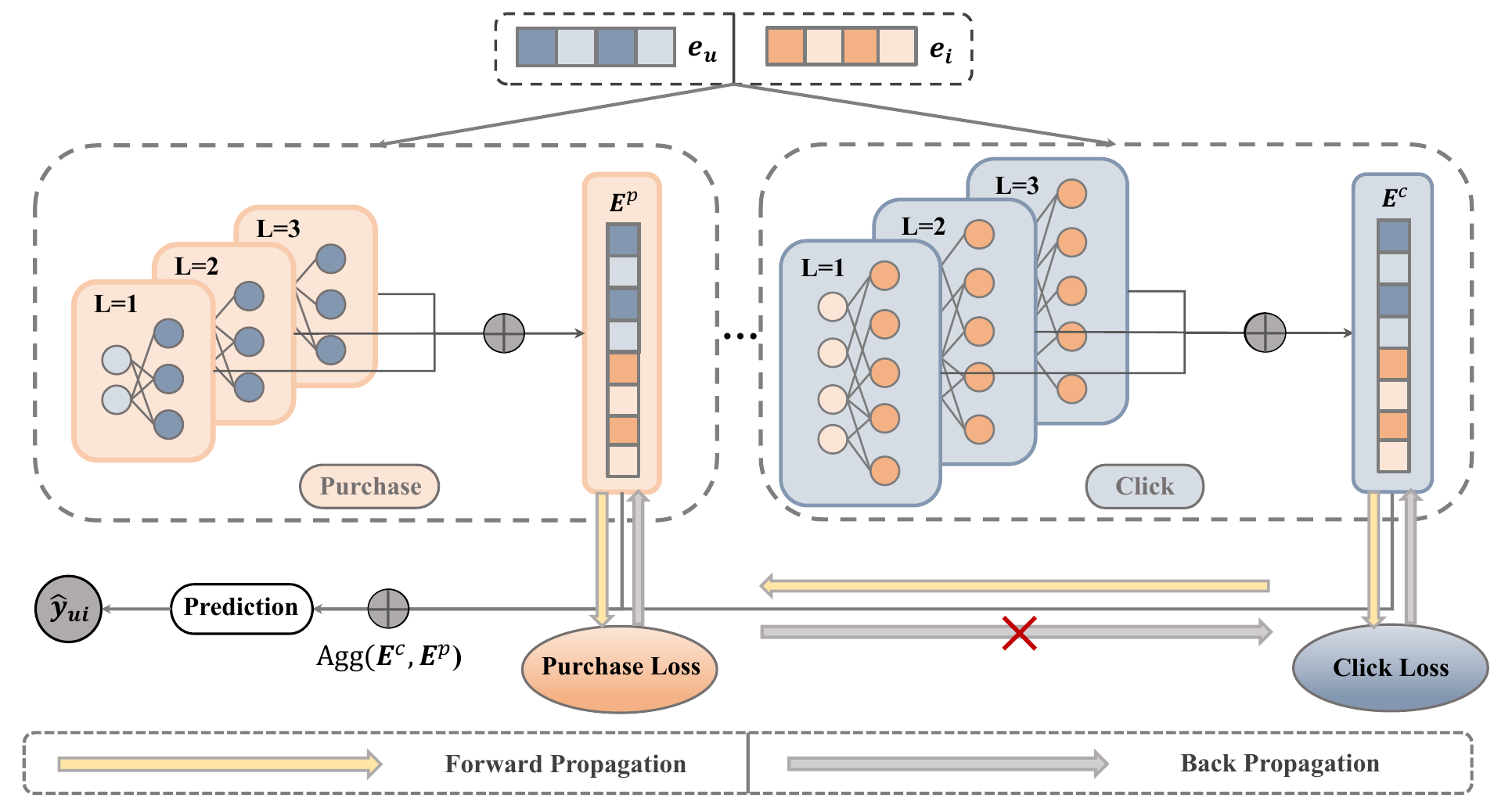}
    \caption{The overall architecture of \model with the example of the \textit{click} behavior graph with rich scale and the \textit{purchase} behavior graph with sparse scale.}
    \vspace{-0.8em}
    \label{fig:framework}
\end{figure*}

Inspired by ESSM~\cite{ma2018entire}, we can regard the interaction prediction of each behavior as a task and learn comprehensive representations from all behavior data by multi-task learning (MTL)~\cite{vandenhende2021multi}. Specifically, we unify all the tasks and update the model parameters to acquire comprehensive representations, which can be defined as:
\begin{align}
     \mathcal{L} = \sum_{k=1}^{K} \mathcal{L}_k(\mathcal{U}, \mathcal{I}, \bm{R}^k),
\end{align}
where $\mathcal{L}_k$ is the loss function of task $k$ conducted on the $k$-th behavior domain. 

\subsection{Graph Collaborative Filtering}
In graph collaborative filtering (GCF) with GNN, the interaction matrix $R^k$ for behavior domain $k$ is employed to construct the graph information $\mathcal{G}^k=(\mathcal{V}, \mathcal{E}^k)$. The core of the GCF model is to recursively aggregate information from the neighborhood to the center node. A single GCN layer~\cite{kipf2016semi} for GCF can be defined as:
\begin{align}
    {\bm{H}^k}^{(l+1)} = \sigma(\Tilde{\bm{A}}^k {\bm{H}^k}^{(l)} {\bm{W}^k}^{(l)}),
\end{align}
where $\Tilde{\bm{A}}^k={\bm{D}^k}^{-\frac{1}{2}} \bm{A}^k {\bm{D}^k}^{-\frac{1}{2}}$ is the laplacian regularized from $\bm{A}^k \in \mathbb{R}^{(M+N)\times(M+N)}$, and $\bm{D}^k \in \mathbb{R}^{(M+N)\times(M+N)}$ is the degree matrix, which is defined as $\bm{D}_{ii}^k=\sum_{j}\bm{A}^k_{ij}$. $W^k \in \mathbb{R}^{d \times d }$ is the learnable parameter, ${\bm{H}^k}^{(l)} \in \mathbb{R}^{(M+N)\times d}$ is the $l$-th layer graph embeddings of behavior domain $k$, and $\sigma(\cdot)$ is the activation function.

Specifically, we adopt LightGCN~\cite{he2020lightgcn} as the GCF backbone for \model due to its wide use for GCF and efficiency. The LightGCN layer for single-behavior domain $k$ is:
\begin{align}\label{eq.3}
     {\bm{H}^k}^{(l+1)} = \Tilde{\bm{A}}^k {\bm{H}^k}^{(l)}.
\end{align}

After propagating $L$ layers, we can obtain the matrices $[{\bm{H}^{k}}^{(0)}, \dots, {\bm{H}^k}^{(L)}]$, which are the representations of $l$-hop neighborhood nodes in behavior domain $k$. Following the setting of previous GCF works, we aggregate all layers of graph embeddings to predict the interaction probability between users and items. Generally, the probability can be calculated by dot product~\cite{koren2009matrix} or multiple layer perceptron (MLP)~\cite{wang2019neural}:
\begin{align}
 \hat{y}_{ui}^k={\bm{e}_u^k}^\top \cdot \bm{e}_i^k \quad \mathrm{or} \quad \hat{y}_{ui}^k=\mathrm{MLP}(\bm{e}_u^k, \bm{e}_i^k),
\end{align}
where $\hat{y}_{ui}^k$ is the predicted interaction probability of user $u$ and item $i$ is the $k$-th behavior graph.

\section{Methodology}
In this section, we present our proposed \model in detail. The overall framework of \model is depicted in Figure~\ref{fig:framework}, which contains three main steps: 
(i) Firstly, \model encodes representations of users and items in different behavior graphs with the GNN-based collaborative filtering model under a multi-task learning architecture, in which learning on each behavior domain indicates a training task.
(ii) Secondly, to alleviate the scale imbalance of different types of behaviors, we enhance the representation learning over the sparse behavior graph (the \textit{purchase} behavior in Figure~\ref{fig:framework}) by aggregating the representation from the behavior graph with a rich scale (the \textit{click} behavior in Figure~\ref{fig:framework}).
(iii) Finally, \model improves the target behavior conversion of users with joint optimization of the multi-behavior learning tasks.

\subsection{Multi-Behavior Graph Collaborative Filtering}
Following the setting of general graph collaborative filtering~\cite{he2020lightgcn,chen2022neighbor}, we let $\bm{e}_u, \bm{e}_i \in \mathbb{R}^d$ as the embedding of the user $u$ and the item $i$, where $d$ is the embedding dimension.
The overall node embedding matrix with $M$ users and $N$ items is denoted as $\bm{E} = [\bm{e}_{u_1}; \dots; \bm{e}_{u_M}; \bm{e}_{i_1}, \dots; \bm{e}_{i_N}] \in \mathbb{R}^{(M + N)\times d}$.
Then, for the specific behavior $k$, the adjacent matrix $\bm{A}^k$ can be defined as:
$$\bm{A}^k = \left[
\begin{array}{cc}
    0 & \bm{R}^k \\
    {\bm{R}^k}^\top & 0
\end{array}
\right].
$$


We then adopt LightGCN~\cite{he2020lightgcn} as the graph collaborative filtering backbone to propagate the information of neighborhood nodes, which can rewrite Eq.(\ref{eq.3}) as follows:
\begin{align}
    {\bm{E}^k}^{(l+1)}=\Tilde{\bm{A}}^k {\bm{E}^k}^{(l)},
\end{align}
where $\Tilde{\bm{A}}^k$ is the laplacian regularized matrix of $\bm{A}^k$, and the ${\bm{E}^k}^{(0)} = \bm{E}$.

After propagating with $L$ layers, we obtain $L$ graph representations $[{\bm{E}^k}^{(0)}, \dots, {\bm{E}^k}^{(l)}]$. Thus, we aggregate them to obtain the final graph representation for behavior $k$:
\begin{align}
    \bm{E}^k=\frac{1}{L+1}\sum_{l=0}^L {\bm{E}^k}^{(l)},
\end{align}
where $1/(L + 1)$ is a uniform weight set to each layer of obtained embedding.

\subsection{Sparse Behavior Enhancement}
Existing methods mainly ignore the consideration of the data imbalance problem between different behaviors~\cite{wang2023multi} , commonly assuming that the availability of balanced training data for different behavioral tasks is ensured through the incorporation of negative samples. 
For instance, in the training process of MMOE~\cite{ma2018modeling}, although interactive data of different behaviors exhibits an imbalance phenomenon (positive samples), it can feed both the interacted positive samples and uninteracted negative samples simultaneously of each task into the DNN for model training. However, as we know, positive samples are far more significant for user modeling due to the uninteracted negative samples do not represent users' negative feedback~\cite{xie2021deep}. Furthermore, for graph collaborative filtering, only the interacted samples are utilized for message propagation.

Therefore, we design an aggregation enhancement for the sparse behavior domain graph representation learning that can leverage the behavior domain $r$ with rich scale to the behavior domain $s$ with sparse scale, to alleviate the data imbalance problem.
Specifically, for a given behavior domain $s$ with sparse scale, we enhance its collaborative filtering with the abundant information learned for the behavior domain $r$ with rich scale, which can be defined as:
\begin{equation}
    {\bm{E}^{s}}^*=\mathrm{Agg}(\bm{E}^{s}, \bm{E}^{r}),
    \label{eq:agg}
\end{equation}
where ${\bm{E}^{s}}^*$ represents the aggregated representation of sparse behavior $s$, $\bm{E}^{s}$ and $\bm{E}^{r}$ are the embeddings obtained by GCF from sparse behavior $s$ and rich behavior $r$, respectively. We utilize the aggregated representation ${\bm{E}^{s}}^*$ for the task prediction of behavior $s$. $Agg(\cdot)$ is the aggregation function where we adopt the mean pooling or concatenation operator for simplicity in this paper. We compare the performance of these two operators in the ablation study.

\subsection{Multi-Task Joint Optimization with Data Imbalance}
Inspired by the multi-task learning paradigm~\cite{ma2018modeling,zhang2021survey}, we treat each behavior prediction as a task and optimize them under the multi-task learning architecture. 
We adopt the BPR loss~\cite{rendle2012bpr} to train the multiple tasks of multi-behavior, which can be formulated as follows:
\begin{gather}
    \mathcal{L}^k = -\sum_{(u,i) \in {\mathcal{O}^k}^+,(u, j) \notin  {\mathcal{O}^k}^+} \ln \sigma (\hat{y}_{ui} ^ k - \hat{y}_{uj}^k),
    \\
    \mathcal{L} = \sum_{k=1}^{K} \mathcal{L}^k,
\end{gather}
where $\sigma(\cdot)$ is the sigmoid function, $\hat{y}_{ui}^k={\bm{e}_u^k}^\top \cdot \bm{e}_i^ k$ and ${\mathcal{O}^k}^+=\{(u,i)|\bm{R}_{u,i}^ k=1\}$ denotes the interaction pair set of users and items in the $k$-th behavior domain. $\mathcal{L}$ is the overall training loss, which is optimized by the gradient descent algorithm, and $\bm{e}_u^k$ and $\bm{e}_i^k$ belong to $\bm{E}^k$, which is obtained after the aggregation if $k$ is the sparse behavior domain.

Furthermore, to prevent the negative impact of sparse behavior domain $s$ on the rich behavior $r$, during the training process, we inhibit the optimization of parameters for abundant data tasks by modifying the gradient descent path as follows:
\begin{align}
    {\bm{E}^{s}}^{'} = \bm{E}^{s} - \eta \frac{\partial{\mathcal{L}^{s}}}{\partial{\bm{E}^{s}}},
\end{align}
where $\eta$ is the learning rate and $\bm{E}^{s}$ is the embedding matrix of behavior domain $s$ by graph collaborative filtering before the representation aggregation with the rich behavior domain, rather than the utilization of $\bm{E}^{s*}$.

\section{Experiment}
\subsection{Experimental Setup}
\subsubsection{\textbf{Datasets}}
We adopt two real-world multi-behavior datasets to present the target behavior conversion performance of \model: Beibei\footnote{\url{https://github.com/Sunscreen123/Beibei-dataset}}~\cite{gao2019neural} and Taobao\footnote{\url{https://tianchi.aliyun.com/dataset/dataDetail?dataId=649}}~\cite{zhu2018learning}.
All datasets catalog three types of user behaviors: click, cart, and purchase.
The dataset preprocessing and splitting follow the widely adopted in previous graph collaborative filtering works~\cite{wang2019neural,he2020lightgcn}, which filter out cold-start nodes with less than 20 interactions and randomly select 80\% of historical interactions as training data and the remaining 20\% as testing data. The detailed statistics of these datasets are illustrated in Table~\ref{tab:stats}.

\begin{table}[htbp]
  \centering
  \small
  \caption{Statistics of the experimental datasets.}
  \vspace{-1em}
  \resizebox{\linewidth}{!}{
    \begin{tabular}{c|c|c|c|c|c}
    \toprule
    Dataset & \# Users & \# Items & \# Click & \# Cart & \# Purchase \\
    \midrule
    Beibei & 21,716 & 7,977 & 2,412,586 & 642,622 & 304,576 \\
    Taobao & 9,773 & 11,268 & 389,632 & 19,145 & 22,115 \\
    \bottomrule
    \end{tabular}%
    \vspace{-1em}
  }
  \label{tab:stats}%
\end{table}%

\subsubsection{\textbf{Baselines}}
To evaluate the effectiveness of our model, we compare \model with three representative models related to our work. For single behavior learning methods, we include MF-based model MF-BPR~\cite{rendle2012bpr} and GCF-based model LightGCN~\cite{he2020lightgcn}. For the multi-task learning method, we include MMOE~\cite{ma2018modeling} with the input embedding learned by LightGCN.

\subsubsection{\textbf{Parameter Settings \& Evaluation Metrics}}
Regarding the model training, we set the embedding size as 64, the batch size as 2048, and the learning rate as 0.001. The models are optimized with tthe Adam optimizer~\cite{kingma2014adam}. To evaluate the top-$K$ recommendation performance on the target \textit{purchase} behavior prediction, we adopt two widely-used evaluation metrics, including Recall@$K$ and NDCG@$K$ ($K$=20).

\begin{table}[tbp]
  \centering
  \caption{Overall comparison results of the target \textit{purchase} prediction.}
  \resizebox{\linewidth}{!}{
    \begin{tabular}{c|c|c|c|c}
    \toprule
    \multirow{2}[4]{*}{Model} & \multicolumn{2}{c|}{Beibei} & \multicolumn{2}{c}{Taobao} \\
\cmidrule{2-5}    \multicolumn{1}{c|}{} & \multicolumn{1}{p{5em}|}{Recall@20} & \multicolumn{1}{c|}{NDCG@20} & \multicolumn{1}{c|}{Recall@20} & \multicolumn{1}{c}{NDCG@20} \\
    \midrule
    MF-BPR    & 0.1004  & 0.0733  & 0.0021  & 0.0014  \\
    LightGCN & 0.1250  & 0.0938  & 0.0125  & 0.0065  \\
    \midrule
    MMOE & \underline{0.1261} & \underline{0.0943} & \underline{0.0226} & \underline{0.0110} \\
    \midrule
    \model (Ours)  & \textbf{0.1549}  & \textbf{0.1194}  & \textbf{0.0307}  & \textbf{0.0164}  \\
    \textit{Improv.} (\%) & 22.84\% & 26.62\% & 35.84\% & 49.09\% \\
    \bottomrule
    \end{tabular}%
    }
  \label{tab:result}%
\end{table}%

\begin{table}[t]
  \centering
  \caption{Ablation study results on \model variants.}
  \resizebox{\linewidth}{!}{
    \begin{tabular}{c|c|c|c|c}
    \toprule
    \multirow{2}[4]{*}{Variant} & \multicolumn{2}{c|}{Beibei} & \multicolumn{2}{c}{Taobao} \\
\cmidrule{2-5}          & Recall@20 & NDCG@20 & Recall@20 & NDCG@20 \\
    \midrule
    \model-\textit{w/o SBE} & 0.1464 & 0.1111 &    0.0207   & 0.0090 \\
    \midrule
    \model-\textit{Concat} & 0.1472 & 0.1129 & \textbf{0.0445} & \textbf{0.0220} \\
    \model-\textit{Mean} & \textbf{0.1549} &  \textbf{0.1194} & 0.0307 & 0.0164 \\
    \bottomrule
    \end{tabular}%
    }
  \label{tab:ablation}%
\end{table}%

\subsection{Main Results}
The overall comparison results are showcased in Table~\ref{tab:result}. The improvement is calculated by comparing \model with the best-performed baseline.
From the comparison results, we can have the following observations:
(i) Compared with MMOE and the single behavior learning models, we can find that the consideration of multiple behaviors as multi-task learning is meaningful for graph collaborative filtering models.
(ii) Our proposed \model outperforms the baseline models with significant improvement on all the experimental datasets.
Specifically, the average improvement on Recall@20 and NDCG@20 of the experimental datasets is 29.34\% and 37.85\%, respectively. The superior performance verifies the effectiveness of our proposed \model.

\subsection{Ablation Study}
Table~\ref{tab:ablation} reports the ablation study results of \model and its ablation variants, where the \model-\textit{w/o SBE} variant removes the aggregation step for sparse behavior representation enhancement, and the \model-\textit{Concat} and \model-\textit{Mean} variants equip the aggregation function of sparse behavior enhancement in Eq.(\ref{eq:agg}) with concatenation and mean pooling operations, respectively.
The results show that \model outperforms its variant without sparse behavior enhancement, which demonstrates the designed component of \model is helpful for multi-behavior graph collaborative filtering.
Further, simply parameter-free concatenation or mean pooling aggregation can sufficiently help to enhance the model performance.

\section{Conclusion}
In this paper, we propose \model, a simple but effective method for multi-behavior graph collaborative filtering to alleviate the behavior scale imbalance problem. Specifically, \model first learns multiple behavior graphs with multi-task graph collaborative filtering architecture.
Then, \model enhances the training over sparse behavior graph by aggregating the learned representation on the behavior graph with more rich interactions. 
Finally, \model joint optimizes the multiple learning tasks to improve the target behavior task conversion performance.
Extensive experimental results on two widely used multi-behavior benchmark datasets illustrate the effectiveness of our proposed model.
In future works, we will explore the enhancement of \model with the consideration of more complex interrelationships between different behaviors.


\bibliographystyle{IEEEtran}
\bibliography{IEEEabrv,reference}

\end{document}